\documentstyle[12pt,epsfig]{article}
\begin{document}
\renewcommand{\theequation}{\thesection.\arabic{equation}}
\newcommand{\eqn}[1]{(\ref{#1})}
\renewcommand{\section}[1]{\addtocounter{section}{1}
\vspace{5mm} \par \noindent
  {\large \bf \thesection . #1}\setcounter{subsection}{0}
  \par
   \vspace{2mm} } 
\newcommand{\sectionsub}[1]{\addtocounter{section}{1}
\vspace{5mm} \par \noindent
  {\bf \thesection . #1}\setcounter{subsection}{0}\par}
\renewcommand{\subsection}[1]{\addtocounter{subsection}{1}
\vspace{2.5mm}\par\noindent {\bf  \thesubsection . #1}\par
 \vspace{0.5mm} }
\renewcommand{\thebibliography}[1]{ {\vspace{5mm}\par \noindent{\bf
References}\par \vspace{2mm}}
\list
 {\arabic{enumi}.}{\settowidth\labelwidth{[#1]}\leftmargin\labelwidth
 \advance\leftmargin\labelsep\addtolength{\topsep}{-4em}
 \usecounter{enumi}}
 \def\newblock{\hskip .11em plus .33em minus .07em}
 \sloppy\clubpenalty4000\widowpenalty4000
 \sfcode`\.=1000\relax \setlength{\itemsep}{-0.4em}}
\def\a{& \hspace{-5pt}}
\def\bea{\begin{eqnarray}}
\def\eea{\end{eqnarray}}
\def\be{\begin{equation}}
\def\ee{\end{equation}}
\def\alp{\alpha}
\def\bet{\beta}
\def\gam{\gamma}
\def\del{\delta}
\def\eps{\epsilon}
\def\sig{\sigma}
\def\lam{\lambda}
\def\Lam{\Lambda}
\def\m{\mu}
\def\n{\nu}
\def\r{\rho}
\def\s{\sigma}
\def\d{\delta}
\newcommand\rf[1]{(\ref{#1})}
\def\nn{\nonumber}
\newcommand{\sect}[1]{\setcounter{equation}{0} \section{#1}}
\renewcommand{\theequation}{\thesection .\arabic{equation}}
\newcommand{\NPB}[3]{{Nucl.\ Phys.} {\bf B#1} (#2) #3}
\newcommand{\CMP}[3]{{Commun.\ Math.\ Phys.} {\bf #1} (#2) #3}
\newcommand{\PRD}[3]{{Phys.\ Rev.} {\bf D#1} (#2) #3}
\newcommand{\PLB}[3]{{Phys.\ Lett.} {\bf B#1} (#2) #3}
\newcommand{\JHEP}[3]{{JHEP} {\bf #1} (#2) #3}
\newcommand{\ft}[2]{{\textstyle\frac{#1}{#2}}\,}
\def\e{\epsilon}
\def\st{\scriptstyle}
\def\mco{\multicolumn}
\def\epp{\epsilon^{\prime}}
\def\vep{\varepsilon}
\def\ra{\rightarrow}
\def\ab{\bar{\alpha}}
\newcommand{\dt}{\partial_{\langle T\rangle}}
\newcommand{\dtbar}{\partial_{\langle\bar{T}\rangle}}
\newcommand{\al}{\alpha^{\prime}}
\newcommand{\mst}{M_{\scriptscriptstyle \!S}}
\newcommand{\mpl}{M_{\scriptscriptstyle \!P}}
\newcommand{\dv}{\int{\rm d}^4x\sqrt{g}}
\newcommand{\lv}{\left\langle}
\newcommand{\rv}{\right\rangle}
\newcommand{\ph}{\varphi}
\newcommand{\sbar}{\,\bar{\! S}}
\newcommand{\xbar}{\,\bar{\! X}}
\newcommand{\fbar}{\,\bar{\! F}}
\newcommand{\zbar}{\,\bar{\! Z}}
\newcommand{\tbar}{\bar{T}}
\newcommand{\ubar}{\bar{U}}
\newcommand{\ybar}{\bar{Y}}
\newcommand{\phb}{\bar{\varphi}}
\newcommand{\cm}{Commun.\ Math.\ Phys.~}
\newcommand{\pr}{Phys.\ Rev.\ D~}
\newcommand{\prl}{Phys.\ Rev.\ Lett.~}
\newcommand{\pl}{Phys.\ Lett.\ B~}
\newcommand{\ibar}{\bar{\imath}}
\newcommand{\jbar}{\bar{\jmath}}
\newcommand{\np}{Nucl.\ Phys.\ B~}
\newcommand{\gsi}{\,\raisebox{-0.13cm}{$\stackrel{\textstyle
>}{\textstyle\sim}$}\,}
\newcommand{\lsi}{\,\raisebox{-0.13cm}{$\stackrel{\textstyle
<}{\textstyle\sim}$}\,}

\thispagestyle{empty}

\begin{center}
\hfill SPIN-1998/13\\
\hfill LMU-TPW 98-17\\[3mm]
\hfill{\tt hep-th/9812071}\\

\vspace{2cm}

{\Large\bf Anomalous couplings for D-branes and O-planes}\\[3mm]
\vspace{1.4cm}
{\sc Jose F. Morales$^{a,e}$, Claudio A. Scrucca$^{b,e}$ and Marco
Serone$^c$} \\
\vspace{1.3cm}

${}^a${\em INFN - Section of Rome Tor Vergata}\\
{\em Department of Physics, University of Rome Tor Vergata}\\
{\em Via della Ricerca Scientifica 1, 00133 Rome, Italy}\\
{\footnotesize \tt morales@roma2.infn.it}\\

\vspace{.5cm}

${}^b$
{\em Department of Physics, Ludwig Maximilian University of Munich}\\
{\em Theresienstra\ss e 37, 80333 Munich, Germany}\\
{\footnotesize \tt Claudio.Scrucca@physik.uni-muenchen.de}

\vspace{.5cm}

${}^c$
{\em Department of Mathematics, University of Amsterdam}\\
{\em Plantage Muidergracht 24, 1018 TV Amsterdam} \\
\& \\ {\em Spinoza Institute, University of Utrecht} \\
{\em Leuvenlaan 4, 3584 CE Utrecht, The Netherlands} \\
{\footnotesize\tt serone@wins.uva.nl}\\

\end{center}

\vspace{0.8cm}

\centerline{\bf Abstract}
\vspace{2 mm}  
\begin{quote}\small
We study anomalous Wess-Zumino couplings of D-branes and O-planes
in a general background and derive them from a direct string computation
by factorizing in the RR channel various one-loop amplitues. In particular,
we find that Op-planes present gravitational anomalous couplings involving
the Hirzebruch polynomial $\hat {\cal L}$, similarly to the roof genus $\hat {\cal A}$
encoding Dp-brane anomalous couplings. We determine, in each case, the precise 
dependence of these couplings on the curvature of the tangent and normal bundles.

\end{quote}
\vfill
\begin{flushleft}
\rule{16.1cm}{0.2mm}\\[-3mm]
\end{flushleft}
{\footnotesize ${}^e$ On leave from {\it SISSA/ISAS, via Beirut 2, 
34100 Trieste, Italy}}
\newpage
\setcounter{equation}{0}


\section{Introduction}

It is known from the work of several authors \cite{pol,dou,li,bsv,ghm,cy,mm} 
(see \cite{bachlec} for a review) that the Wess-Zumino coupling in the effective 
world-volume theory of a Dp-brane in a generic supergravity background takes the 
following form:
\be
S=\mu_{p} \int \,C\wedge\, e^{{\cal F}}\, \wedge \, \left.
\sqrt{\hat{{\cal A}}({\cal R_T})/\hat{{\cal A}}({\cal R_N})}\right|_{(p+1)-form}
\label{WZD}
\ee
where $C=\sum_n C_{(n)}$ is the formal sum over all Ramond Ramond (RR)
form potentials,
pulled back to the world-volume of the D-brane, ${\cal F} = 2 \pi \al F - B$ 
is the gauge-invariant combination of the field strength $F$ of the gauge
field living on the D-brane and the pull-back of the Neveu-Schwarz 
Neveu-Schwarz (NSNS) antisymmetric 
tensor field $B$. ${\cal R_T}$ and ${\cal R_N}$ are the tangent and normal 
components of the appropriately normalized curvature two-form 
${\cal R} = 4 \pi^2 \al R$ on the world-volume, and $\hat {\cal A}({\cal R_T})$ and 
$\hat {\cal A}({\cal R_N})$ are the {\it roof genus polynomials} of the tangent and 
normal bundle of the Dp-brane. 
$\hat {\cal A}({\cal R})$ is a polynomial of the curvature two-form, and
in terms of Pontrjagin classes $p_n(R)$ one has
\be
\sqrt{\hat {\cal A}({\cal R})} = 1 - \frac {(4 \pi^2 \al)^2}{48} p_1(R) 
+ \frac {(4 \pi^2 \al)^4}{2560} {p_1}^2(R) - \frac {(4 \pi^2 \al)^4}{2880} p_2(R) + ...
\ee
The notation of (\ref{WZD}) is then clear: in expanding all the forms one has 
to pick up only the (p+1)-form integrated over the (p+1)-dimensional world-volume.
All the terms appearing in (\ref{WZD}), but the usual minimal coupling to
$C_{(p+1)}$, can be related to gauge and gravitational anomalies arising
in the world-volume theories of certain brane intersections containing chiral 
fermions  \cite{ghm}.
In such cases one can get rid of the anomaly appearing on the world-volume 
theory by a bulk term, with support only on the world-volume, that cancels it.
This is basically how the inflow-mechanism works \cite{ch}.
These anomaly considerations have been actually crucial to establish the existence
of most of the terms appearing in (\ref{WZD}), that 
are then called {\it anomalous couplings}. Anyway, although the couplings induced
by the field strenght ${\cal F}$ have been confirmed in various ways,
also with explicit computations \cite{li}, the gravitational ones coming from 
$\hat{{\cal A}}({\cal R_T})$ and $\hat{{\cal A}}({\cal R_N})$ are more difficult
to analyze.
Up to now several checks have been performed \cite{bsv,cr}, but a direct 
computation confirming the presence of $\hat{{\cal A}}({\cal R_N})$ and the 
eight-form term in the expansion of $\hat{{\cal A}}({\cal R_T})$\footnote{This is, 
of course, the last term that can appear in the expansion of the roof genus in 
(\ref{WZD}).} has not been performed yet.

Similarly to D-branes, also orientifold planes have anomalous couplings 
beside the minimal coupling to RR-forms. As argued in \cite{dasjatmu,dasmu}, 
these are again required to cancel anomalies in chiral world-volume field theories, 
the novel feature being the appearence of chiral antisymmetric tensors, i.e.
antisymmetric tensors with (anti) self-dual field strenghts, which contribute to
the anomaly beside chiral fermions. Since open strings cannot end on O-planes, 
these do not support world-volume gauge fields, and correspondingly 
only gravitational anomalous couplings occure. We claim that the complete 
Wess-Zumino coupling for an Op-plane is
\be
S=\mu^\prime_{(p)} \int \,C \wedge \, \left.
\sqrt{\hat{{\cal L}}({\cal R_T}/4)/\hat{{\cal L}}({\cal R_N}/4)} \right|_{(p+1)-form}
\label{WZO}
\ee
where $\mu_p^\prime = - 2^{p-4} \mu_p$ is the charge of an O-plane with the 
conventions of \cite{dab} and $\hat {\cal L}({\cal R_T}/4)$ and 
$\hat {\cal L}({\cal R_N}/4)$ are 
the {\it Hirzebruch polynomials} of the tangent and normal bundles of the Op-plane. 
Similarly to before, $\hat {\cal L}({\cal R}/4)$ is a polynomial of the curvature 
2-form $R$, and in terms of Pontrjagin classes $p_n(R)$ one has
\be
\sqrt{\hat {\cal L}({\cal R}/4)} = 1 + \frac {(4 \pi^2 \al)^2}{96} p_1(R) 
- \frac {(4 \pi^2 \al)^4}{10240} {p_1}^2(R) + \frac {7(4 \pi^2 \al)^4}{23040} p_2(R) + ...
\label{sqrtL}
\ee
The dependence on the tangent bundle curvature in (\ref{WZO}) has been already studied 
in \cite{dasjatmu,dasmu}, where the coefficients of the terms quadratic and quartic in 
the curvature were determined. Whereas the quadratic term appearing in the expansion 
(\ref{sqrtL}) is consistent with the results of \cite{dasjatmu,dasmu}, the quartic term shows
a discrepancy. Our results come from a direct string computation and pass various
consistency checks, as will be discussed in detail in the following. We think that 
this gives strong evidence for the correctness of the coefficients appering in 
(\ref{sqrtL})\footnote{We are grateful to S. Mukhi for a useful discussion about this issue.}.
On the other hand, we are not aware of any prior discussion about the anomalous couplings
for Op-planes arising from the normal bundle curvature.

Aim of this work is then to extract the couplings (\ref{WZD}), (\ref{WZO}) from a 
one-loop computation representing the interaction between parallel D-branes and 
O-planes. More precisely, we will be interested in the interaction due to the 
exchange of RR forms in the closed string channel, in the presence of
electromagnetic and gravitational backgrounds. Due to the GSO projections, this 
interaction further splits into two contributions corresponding to the RR even and odd spin 
structures, encoding respectively electric and magnetic interactions \cite{bis}.
From the open string point of view, they correspond to the $(-)^F$ part of the 
partition function in the NS and R sectors. 
Generalizing Polchinski's computation of D-brane charges \cite{pol}, 
one could in principle use equivalently one or the other of these two kinds of RR 
interactions to deduce the couplings. However, in the case at hand the nature of 
the couplings to probe lead to several difficulties in the analysis of the electric 
interaction. Indeed, this interaction can not be computed exactly in the background 
fields. On the other hand, all the difficulties disappear for the
magnetic interaction, that as we shall see is topological and can be computed exactly,
at one-loop level. Moreover, from the open string point of view, the odd spin structure 
represents precisely the anomalous part of a loop of massless chiral particles, and yields
therefore automatically the anomaly related by the inflow mechanism to the
couplings (\ref{WZD}), (\ref{WZO}). 
The direct derivation of the tangent bundle part of the anomalous couplings (\ref{WZD}) and 
(\ref{WZO}) that we present is rigorous and supported by a strong consistency check. 
However, the derivation of the normal bundle part of these couplings suffers,
strictly speaking, of an overall normalization ambiguity, since it relies on an 
analysis of formally vanishing amplitudes, that is not supported by a consistency
check. Nonetheless, we reach a clear understanding of the mechanism
responsible for the different dependence on the tangent and normal bundle curvatures, 
and we therefore believe that our arguments give really strong evidence
for the results we propose.

The plan of the paper is as follows. In sections two and three we consider 
particular one-loop correlation functions on the annulus, M\"obius strip and 
Klein bottle surfaces, from which we will extract the tangent and normal
bundle contributions of the couplings (\ref{WZD}) and (\ref{WZO}).
In section four we rederive, through a $\sigma$-model approach, the results
obtained in sections two and three, emphasizing their generality and including
also the dependence on ${\cal F}$. In section five, following \cite{dasjatmu,dasmu}, we 
fix and check the normalizations entering in the previous loop computations 
through a consistency check with anomaly cancellation in Type I string theory.
In last section, we give some final comments and conclusions, and in an
appendix some technical details about our computations are reported.


\section{Tangent bundle}
\setcounter{equation}{0}

The most direct string computation that one could imagine to do, in order to extract the
induced RR couplings appearing in (\ref{WZD}) and (\ref{WZO}), is just a correlation 
function on a disk and a cross-cup (with the appropriate Neumann or Dirichlet 
boundary conditions) of a certain number of gravitons and world-volume photons 
with the appropriate RR form.
This approach has been used, for instance, in \cite{cr} to check the first non-trivial
term of the expansion of the gravitational term  associated to the tangent bundle of 
D-branes and O-planes. Using this method, the next term in the expansion
would require to compute a 5-point function on a disk or a cross-cup of four gravitons
with a RR tensor field. By choosing the appropriate polarizations for the four
gravitons, in this way one could eventually derive both the gravitational
couplings associated to the tangent and normal bundle. Such kind of computations
is however quite laborious and awkward. Moreover, it does not really display
the topological nature of these interactions, nor determine in a reliable way all the
coefficients.

On the other hand, one could imagine to generalize Polchinski's 
factorization procedure \cite{pol}, and compute appropriate one-loop 
amplitudes from which one can extract the RR couplings
of D-branes and O-planes. In particular, D-D, D-O and O-O interactions are
encoded in amplitudes involving respectively the annulus, M\"obius strip and
Klein bottle surfaces. The effect of the background is taken into account by inserting 
a certain number of the corresponding vertex operators, depending on the order of the 
coupling to be studied.
It is clear that these one-loop correlation functions can have contributions from 
the couplings (\ref{WZD}) and (\ref{WZO}), in which a closed RR string
state is exchanged between the two sources. By a comparison with the same computation 
in the corresponding low-energy effective theory, one can then extract the couplings 
(\ref{WZD}) and (\ref{WZO}).

As already mentioned in the introduction, there are two kinds of RR interactions 
that can take place in the two RR spin structures: electric ones in the even spin 
structure, in which a given RR form propagates, and magnetic ones in the odd 
spin structure, where the intermediate RR form turns into its 
magnetic dual during the propagation (because of the $\Gamma_{11}$ implementing Hodge 
duality on bispinors). Technically, the presence of the fermionic zero modes 
allow for a non-vanishing correlation only between RR forms which are equal up to the 
Hodge duality. The relevant correlators with one potential and one field strength 
are\footnote{Whereas electric correlations of potentials are most natural, magnetic 
correlations are well defined only if at least one of the potentials is turned to a 
field strength, allowing to implement explicitly Hodge duality.}
\bea
\langle C_{(p)}^{M_1 ... M_{p}} H_{(q+1)}^{N_1 ... N_{q+1}} \rangle_{even} 
\a\sim\a \delta_{p,q} \, \delta^{M_1 ... M_p; [N_1 ... N_q} 
\partial^{N_{q+1}]} \Delta 
\label{propeven}\\
\langle C_{(p)}^{M_1 ... M_p}  H_{(q+1)}^{M_1 ... M_{q+1}} \rangle_{odd} 
\a\sim\a \delta_{p+q,8} \, \epsilon^{M_1 ... M_p N_1 ... N_{q+1} P} 
\partial_P \Delta
\label{propodd}
\eea
where $\Delta$ is the free massless scalar propagator in the transverse directions
and the tensor $\delta^{M_1 ... M_p;N_1 ... N_q}$ denotes $\delta^{M_1 N_1} ... 
\delta^{M_p N_q}$ appropriately antisymmetrized in the $M_i$ and 
$N_i$ indices (for $p=q$). The propagators (\ref{propeven}) and (\ref{propodd}) are 
manifestly related through Hodge duality, $H_{(p)} = {}^* H_{(10-p)}$, reflecting the 
fact that the odd spin structure is defined with a $\Gamma_{11}$ insertion (implementing 
this duality). Here and in the following, capital indices $M,N = 0, ..., 9$ run over all
space-time, whereas Greek indices $\mu,\nu=0,...p$ label Neumann directions and
Latin indices $i,j=p+1,...9$ Dirichlet directions.

As anticipated in the introduction, we shall concentrate on the topological magnetic 
interaction encoded in the odd spin structure. In general, the presence of a gravitational 
background not only induces new anomalous couplings, but also modifies the free 
propagators (\ref{propeven}) and (\ref{propodd}). Therefore, in extracting couplings 
by factorization, one has to carefully disentangle them from corrections to the free 
propagator induced by the background. Whereas this difficulty can be avoided
for the tangent bundle contributions of (\ref{WZD}), (\ref{WZO}) by simply taking
the gravitons propagating in the brane world-volume only, such a simplification
does not take over for the normal bundle couplings. These are indeed clearly
visible only if the gravitons have momenta and polarizations along the
transverse directions as well. It can be easily seen, moreover, that the
magnetic interaction (\ref{propodd}) can take place only when the RR tensor fields turned
on by the two branes cover at least nine of the ten space-time directions.
This is actually the case for D8-branes and O8-planes only.
Otherwise, the ten dimensional epsilon tensor will always be vanishing.

We analyze in this section a particular setting which allows to deduce in a rigorous 
way the tangent bundle part of the gravitational anomalous couplings, both for 
D-branes and O-planes. It relies on the annulus and M\"obius strip amplitudes, giving 
respectively the D8-D8 and D8-O8 interactions. We shall then also present a similar 
analysis for the Klein bottle and show that the result is indeed compatible with the 
interpretation of O8-O8 interaction.
The choice of computing one-loop amplitudes for a given kind
of D-branes/O-planes is dictated by definiteness and to simplify the corresponding
field theory analysis. We would like to stress, however, that all the results below 
can be deduced for any choice of parallel D-branes/O-planes, as will be shown in section 
four.

\subsection{Annulus and M\"obius strip}

\vskip 2pt

We shall focus on magnetic interactions which involve one power of the world-volume 
gauge field and four curvature tensors. These are encoded in a correlation function 
with $1$ photon and $4$ gravitons in the RR odd spin structure. 
This interaction is non-vanishing only between two parallel D8-branes 
or between a D8-brane and an O8-plane. 

Before proceeding with the computation, it is essential to recall that in the odd spin 
structure on the cylinder and the M\"obius strip there is a gravitino zero mode 
which is responsible for the insertion of the sum of the left and right 
supercurrents $T_F + \tilde T_F$\footnote{We use here the formalism introduced 
in \cite{fms}.} and a Killing spinor that requires the total  
superghost charge to be $-1$ (instead of $0$
as in the even spin structures). Taking the photon vertex operator in the $-1$ picture, 
and all the graviton vertex operators in the $0$ picture, we have therefore
to evaluate
\be
I_{\gamma g^4} = \langle (T_F + \tilde T_F) V_\gamma^{(-1)}\,V_g^{(0)}\,V_g^{(0)}\,
V_g^{(0)}\,V_g^{(0)}\rangle
\label{4g} 
\ee
both on the annulus and the M\"obius strip. 

\begin{figure}[h]
\vskip 10pt
\centerline{$\raisebox{22pt}{$a)$}$ \epsfysize = 2cm \epsffile{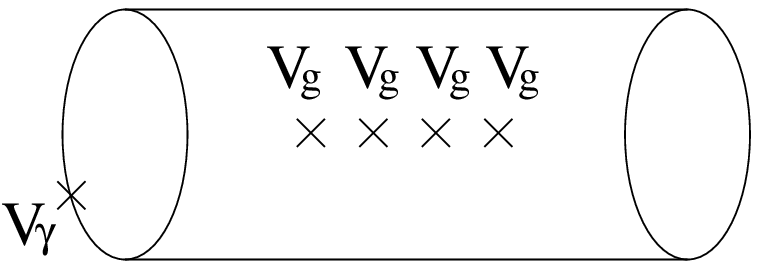}}
\vskip 20pt
\centerline{$\raisebox{22pt}{$b)$}$ \epsfysize = 2cm \epsffile{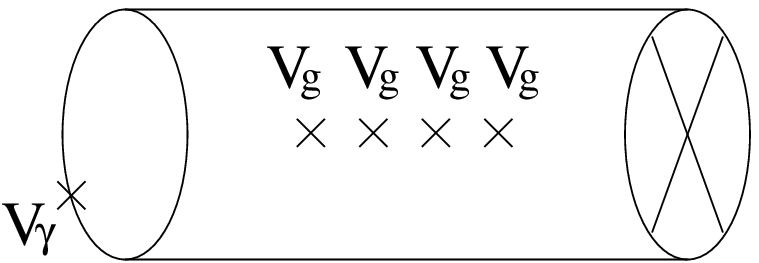}}
\vskip 10pt
\centerline{Fig. 2.1: {\it The correlation $I_{\gamma g^4}$ on a) the annulus and 
b) the M\"obius strip.}}
\vskip 10pt
\end{figure}

The left and right moving picture-changing operators are\footnote{In the following,
we fix $\al = 1/2$. However, we shall not attempt to keep track of the overall normalizations, 
since all the relevant coefficients will be checked in section five.}
\be
T_F = e^\phi \psi^M \partial X_M \;,\;\; 
\tilde T_F = e^{\tilde\phi} \tilde\psi^M \bar \partial X_M
\label{pc}
\ee
In (\ref{4g}), $V_\gamma^{(-1)}$ is the photon vertex operator in the $-1$ picture
\be
V_\gamma^{(-1)}(p)=\oint ds \,e^{-\phi} A_{\mu}(p)\psi^{\mu}e^{ip\cdot X}
\label{vgamma}
\ee
$V_g^{(0)}$ is the usual graviton vertex operator in the 0 picture
\be 
V^{(0)}_g(p)=\int \! d^2z\,h_{MN}(p)
(\partial X^{M}+ip\cdot\psi\psi^{M})
(\bar{\partial}X^{N}+ip\cdot\tilde{\psi}\tilde{\psi}^{N})e^{ip\cdot X} 
\label{vgra} 
\ee
whose leading two-derivative part is
\be
V_g^{lin.}(p)= \frac 12 \int \! d^2z\, R_{MNPQ}(p)
(X^{M}\partial{X}^{N}+\psi^M\psi^N)(X^P\bar{\partial} X^Q +
\tilde{\psi}^P \tilde{\psi}^Q)
\label{vgravlin}
\ee
in terms of the linearized Riemann tensor
\be
R_{MNPQ}(p)=-\frac{1}{2}[p_M\,p_P\,h_{NQ}(p)+
p_N\,p_Q\,h_{MP}(p)-p_N\,p_P\,h_{MQ}(p)-
p_M\,p_Q\,h_{NP}(p)]\! \label{RT}
\ee
For simplicity, we take all the momenta and polarizations of the four gravitons 
in the eight Neumann directions $\mu,\nu=1,..,8$.

In the odd spin structure, the fermion fields $\psi^M$ are completely periodic and 
have therefore zero modes $\psi_0^M$. Correspondingly, the one-loop correlation
function (\ref{4g}) contains an integral over the ten fermionic zero modes 
$\psi_0^M$ which vanishes unless all the ten $\psi_0^M$ are inserted
\be
\int (\prod_{i=1}^{10}d\psi_0^{M_i}) \, \psi_0^{M_1} ... \psi_0^{M_{10}} = 
\epsilon^{M_1 ... M_{10}}
\label{intzm}
\ee
Since $T_F + \tilde T_F$ and $V_\gamma^{(-1)}$ can soak up at most two of them, the 
remaining eight zero modes should be furnished by the gravitons. Although
each graviton vertex contains up to a four fermion term, it is easily seen
that due to the cyclic property of the Riemann tensor, the maximum number of
zero modes that each of them can soak up is actually two. Each graviton has then
to provide precisely two fermionic zero modes, and the vertex (\ref{vgravlin}) is 
replaced by the effective one
\be
\tilde{V}_g^{eff.}=\int \! d^2z\, R_{\mu\nu}(p)\left[
X^\mu(\partial+\bar{\partial})X^\nu+(\psi-\tilde{\psi})^\mu(\psi-\tilde{\psi})^\nu
\right]
\label{veff}
\ee
in terms of the $SO(8)$-valued curvature two-form
\be
R_{\mu\nu}= \frac 12 R_{\mu\nu\rho\sigma} \psi_0^\rho\, \psi_0^\sigma
\ee
The remaining two fermionic zero modes must be provided by the photon vertex and 
the picture changing operator. The latter, in particular, is the only operator that 
can provide the Dirichlet zero mode $\psi_0^9$, whereas the photon vertex (\ref{vgamma}) 
has to provide $\psi_0^0$.
In this way, the correlation (\ref{4g}) factorizes into a longitudinal part involving
only the picture changing and photon vertex operators (and conventionally also
ghosts and superghosts), and a light-cone part involving only the four graviton vertex 
operators. In the first part, the superghost correlation precisely 
cancels the partition function of the two longitudinal fermions along the $0,9$ directions.
Similarly, at leading order in the photon momentum, only the constant mode of 
$\partial_\sigma X^9$, the distance $b$ between the 
two D8 or O8 sources in the nine direction, contribute in $T_F + \tilde T_F$.
The ghost partition function cancels then that of the two bosons along the $0,9$ directions. 
The total longitudinal contribution is then proportional to 
$T (2 \pi t)^{-1/2} e^{- b^2 t/(2 \pi^2)} \, A_0 \, bt$, where $T$ is the total time and $t$ the
modulus of the surface. Equation (\ref{4g}) therefore reduces to 
\be
I_{\gamma g^4} = T \int_0^\infty \! \frac {dt}t \,(2 \pi t)^{-1/2} \,e^{- b^2 t / (2 \pi^2)} 
A_0 \,bt \, I^{eff.}_{g^4}
\ee
in terms of the effective four-graviton correlation function
\be
I^{eff.}_{g^4} = \langle \langle \tilde{V}_g^{eff.}\,\tilde{V}_g^{eff.}\,\tilde{V}_g^{eff.}\,
\tilde{V}_g^{eff.}\,\rangle \rangle
\label{4geff}
\ee
in the eigth space-like world-volume directions.
Surprisingly, the same kind of correlators appeared in a different
context in type I four dimensional compactifications with $N=2$ supersymmetry,
in evaluating a certain class of gravitational couplings, commonly called $F_g$'s
\cite{bcov,ms,mc}. Following \cite{ms,mc}, it is convenient to exponentiate
the correlator (\ref{4geff}), reducing the computation to the evaluation of
the partition function for a twisted action in the RR odd spin structure:
\be
Z(t)=\langle \langle e^{-S_0+S_{int.}} \rangle \rangle
\label{Z}
\ee
where $S_0$ is the free string action and
\be
S_{int}=\int \! d^2z \, R_{\mu\nu}(p)\left[
X^\mu\,(\partial + \bar \partial) X^\nu+(\psi-\tilde{\psi})^\mu(\psi-\tilde{\psi})^\nu
\right]
\label{Sint}
\ee
In operatorial formalism, the odd spin structure partition function $Z(t)$ on the 
annulus and M\"obius strip is defined as a trace over open string states:
\be
Z_A (t) = \mbox{Tr}_{R}[(-)^F e^{- t H}] \;,\;\; 
Z_M (t) = \mbox{Tr}_{R}[(-)^F \Omega e^{- t H}]
\label{ZAM}
\ee
where $H$ is the open string Hamiltonian in a general background and $(-)^F$ and 
the world-sheet parity operator $\Omega$ implement the appropriate boundary 
conditions for bosons and fermions. More precisely, $H$ is the hamiltonian of a 
two-dimensional supersymmetric non-linear $\sigma$-model in a generic 
eight-dimensional target manifold. Since $\Omega$ commutes with the conserved 
linear combination of world-sheet supercharges $Q+\tilde{Q}$, $Z_{A,M} (t)$
are both topological indices \cite{wit}; they do not depend on the modulus $t$ and
can be computed exactly. We will discuss the path-integral representation of 
(\ref{ZAM}) in section four, where we show that it involves
the quadratic interaction (\ref{Sint}). 
Once $Z$ has been evaluated, the four point function (\ref{4geff}) is given by the 
term with four curvatures and eight fermionic zero modes: $I^{eff.}_{g^4} =Z_{TW}|_{g^4}$.
The evaluation of the determinant is straightforward. 
Only the lowest-lying R open string states contribute to the trace, all the others 
cancelling by world-sheet supersymmetry. In terms of the skew-eigenvalues 
$\lambda_i$ of the $R_{\mu\nu}$ matrix, one finds (see section four and appendix)
\be
Z_{A,M} = \int\! d^8x_0 \!\int\! d^8\psi_0 \, 
\prod_{i=1}^4\left(\frac{\lambda_i / 4 \pi}{\sinh \lambda_i /4 \pi}\right)
= \int\! d^8x_0 \int\! d^8\psi_0 \, \hat {\cal A}(R)
\label{ZZAM}
\ee 
Notice that $Z_{A,M}$ are independent of $t$ as expected.

Performing the integral over fermionic zero modes using (\ref{intzm}), the final 
result for both surfaces is the same. 
For the M\"obius strip, it is convenient to factorize an additional $2^{4}$ factor in 
order to recover the correct O8-plane charge $\mu_8^\prime = - 16 \mu_8$. 
This amounts to change $R$ into $R/2$. Integrating over the modulus, the conveniently 
normalized results are then
\bea
I^A_{\gamma g^4} \a=\a T \, \mu_8^2 \int \!d^8x_0 \, \epsilon_{\mu_1 ... \mu_{9}} 
\left(\hat {\cal A}({\cal R})\right)_8^{\mu_1 ... \mu_8}  A^{\mu_9} \partial 
\Delta_{(1)} (b) 
\label{IA}\\
I^M_{\gamma g^4} \a=\a T \, \mu_8 \mu_8^\prime \int \!d^8x_0 \, \epsilon_{\mu_1 ... \mu_{9}} 
\left(\hat {\cal A}({\cal R}/2)\right)_8^{\mu_1 ... \mu_8} A^{\mu_9} \partial
\Delta_{(1)} (b)
\label{IM}
\eea
where $\Delta_{(1)}(b)$ is the scalar Green function in the transverse dimension.

\begin{figure}[h]
\vskip 10pt
\centerline{$\raisebox{22pt}{$a)$}$ \epsfysize = 2cm \epsffile{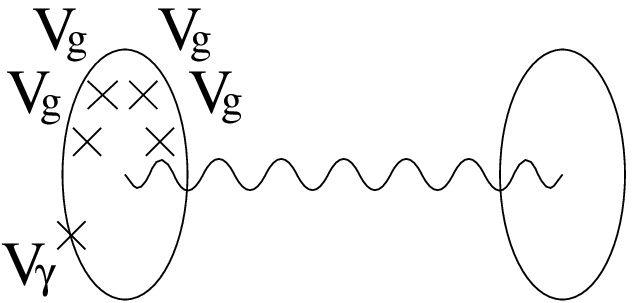} 
$\raisebox{22pt}{$+\!\!$}$ \epsfysize = 2cm \epsffile{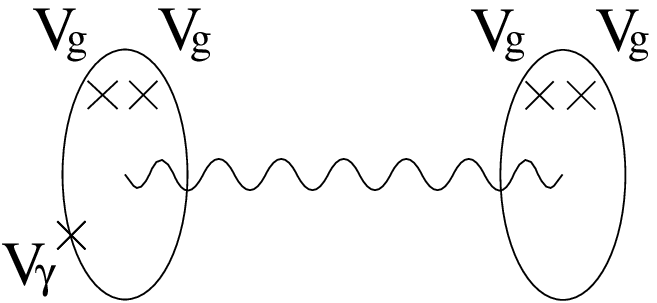} 
$\raisebox{22pt}{$+\!\!$}$ \epsfysize = 2cm \epsffile{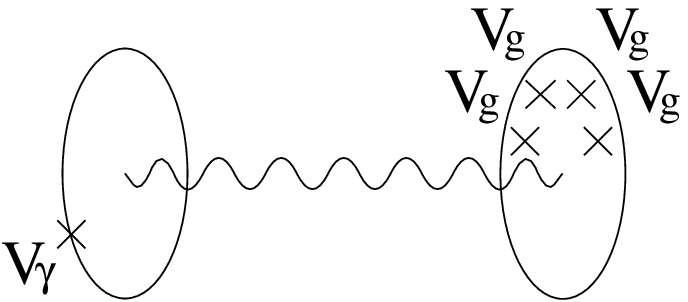}}
\vskip 20pt
\centerline{$\raisebox{22pt}{$b)$}$ \epsfysize = 2cm \epsffile{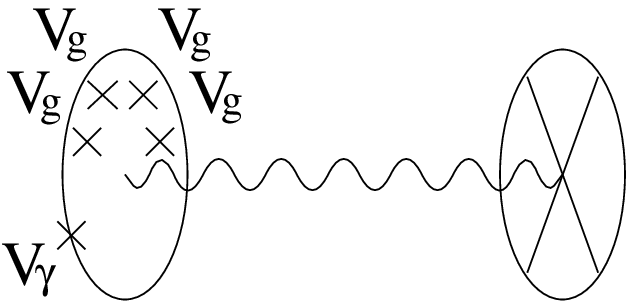} 
$\raisebox{22pt}{$+\!\!$}$ \epsfysize = 2cm \epsffile{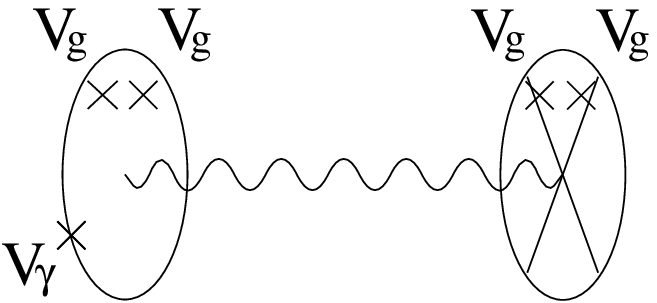} 
$\raisebox{22pt}{$+\!\!$}$ \epsfysize = 2cm \epsffile{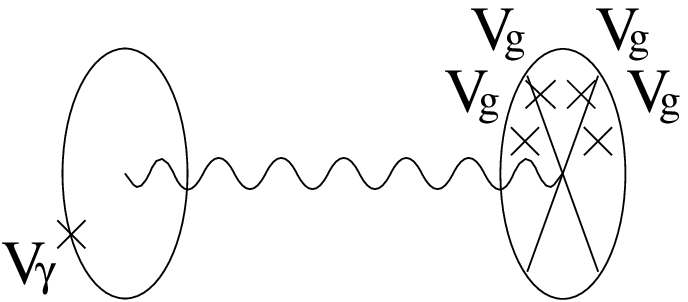}}
\vskip 10pt
\centerline{Fig. 2.2: {\it Factorization of $I_{\gamma g^4}$ on a) the annulus
and b) the M\"obius strip.}}
\vskip 10pt
\end{figure}

It is now straightforward to use (\ref{IA}) and (\ref{IM}) to deduce the 
gravitational anomalous couplings (\ref{WZD}) and (\ref{WZO}).
Consider indeed (\ref{WZD}) and (\ref{WZO}) with some unknown 
curvature polynomial $\hat {\cal D}({\cal R})$ and $\hat {\cal O}({\cal R})$ to be determined.
One of the sources
is always a D8-brane with a world-volume gauge field, for which the part of the 
Wess-Zumino coupling which is linear in the gauge field is ($H = d C$)
\be
S_{D8} (A) = \mu_8 \int \left. A \wedge H \wedge \hat {\cal D}({\cal R}) \right|_{9-form}
\label{CD8A}
\ee
as can be shown by integrating by parts ($\hat {\cal D}({\cal R})$ 
and $\hat {\cal O}({\cal R})$ are closed forms). 
Note that the world-volume gauge field $A$ is crucial to turn the RR tensor 
fields to their field strengths, allowing then to perform explicitly the Hodge duality 
operation leading to the non-vanishing correlations (\ref{propodd}). 
The other source is either a D8-brane without gauge field or a O8-plane,
with Wess-Zumino couplings
\be
S_{D8} = \mu_8 \int \left. C \wedge \hat {\cal D}({\cal R}) \right|_{9-form} \;,\;\;
S_{O8} = \mu^\prime_8 \int \left. C \wedge \hat {\cal O}({\cal R}) \right|_{9-form}
\label{CD8O8}
\ee
It is then easy to compute, using the propagator (\ref{propodd}), 
the corresponding magnetic interactions between two D8-branes, and a D8-brane 
and an O8-plane, as shown in 
figure 2.2\footnote{Similar interactions were already considered in \cite{fer}
for the D0-D8 system.}.
The results are then\footnote{Actually,
to obtain these results one has to consider a $0$-form term $H_{(0)}$ in $H$
of (\ref{CD8A}), which would correspond formally to a $-1$-form $C_{(-1)}$ in $C$. 
This is closely related to the fact that we are
looking at a truncation of the theory, which does not satisfy tadpole cancellation 
\cite{tad,polcai}. Indeed, the unphysical $C_{(9)}$ form, dual to $C_{(-1)}$, that we 
encounter for D8-branes and O8-planes corresponds, by T-duality, to the $C_{(10)}$
form for D9-branes and O9-planes, which lead to inconsistent tadpoles in Type I theory
unless $G=SO(32)$ \cite{polcai}. This is a signal that a IIA background with D8-branes 
only, at least at weak coupling, is inconsistent.}
\bea
I_{D8-D8} \a=\a T \, \mu_8^2 \int \!d^8x_0 \, \epsilon_{\mu_1 ... \mu_{9}} 
(\hat {\cal D}({\cal R}) \wedge \hat {\cal D}({\cal R}))_8^{\mu_1 ... \mu_8} A^{\mu_9} 
\partial \Delta_{(1)} (b) 
\label{ID8D8}\\
I_{D8-O8} \a=\a T \, \mu_8 \mu_8^\prime \int \!d^8x_0 \, \epsilon_{\mu_1 ... \mu_{9}} 
(\hat {\cal D}({\cal R}) \wedge \hat {\cal O}({\cal R}))_8^{\mu_1 ... \mu_8} A^{\mu_9} 
\partial \Delta_{(1)} (b)
\label{ID8O8}
\eea
Comparing with (\ref{IA}) and (\ref{IM}), one finds that
\bea
\hat {\cal D} ({\cal R}) \wedge \hat {\cal D} ({\cal R}) \a=\a \hat {\cal A}({\cal R}) 
\label{constrainta} \\
\hat {\cal D} ({\cal R}) \wedge \hat {\cal O} ({\cal R}) \a=\a \hat {\cal A}({\cal R}/2)
\label{constraintm}
\eea
at least for the 8-form component. It is remarkable that, thanks to the property
\be
\sqrt{\hat {\cal A}({\cal R})} \wedge \sqrt{\hat {\cal L}({\cal R}/4)}  
= \hat {\cal A} ({\cal R}/2)
\ee
following from trigonometric identities, (\ref{constrainta}) and 
(\ref{constraintm}) have the unique solution
\be
\hat {\cal D} ({\cal R}) = \sqrt{\hat {\cal A} ({\cal R})} \;,\;\; 
\hat {\cal O} ({\cal R}) = \sqrt{\hat {\cal L} ({\cal R}/4)} 
\label{sol}
\ee
which satisfies (\ref{constrainta}) and (\ref{constraintm}) for all the component 
forms, and not only for the 8-form part. 

\subsection{Klein bottle}

\vskip 2pt

Although the computation performed in last subsection is actually sufficient in order 
to extract the 
gravitational couplings associated to O-planes, we want to show here a direct computation
in the Klein bottle that will confirm the couplings found. Moreover, it will turn out
to be useful in order to establish a connection between these couplings and anomalies 
associated to tensor fields with (anti)self-dual field strenghts, that we
will discuss in the last section.

Unfortunately, it is not possible to probe the interaction between two O8-planes 
with the same set-up as before. Indeed, a 1-photon 4-graviton correlation function 
makes no sense in this case, since O8-planes do not support world-volume 
gauge fields. 
However, it still makes sense to consider the partition function (\ref{Z}) evaluated
on the Klein bottle. In operatorial formalism, this corresponds to a trace over closed
string states
\be
Z_K(t) = \mbox{Tr}_{RR} [(-)^{F+\tilde F}\,\Omega\,e^{- t H}]
\label{ZK}
\ee
where as before $(-)^{F + \tilde F}$ and the world-sheet parity operator $\Omega$ implement 
the appropriate boundary conditions for bosons and fermions. $H$ is again the 
Hamiltonian of a two-dimensional supersymmetric non-linear $\sigma$-model 
in a generic eight-dimensional target manifold. As before, $\Omega$ commutes 
with the conserved linear combination of world-sheet supercharges $Q+\tilde{Q}$, and
the trace above is an index \cite{wit}, independent on the parameter $t$. 
Again, the path-integral representation of  (\ref{ZK}) involves, after expanding in 
normal coordinates,  the quadratic interaction (\ref{Sint}).

The evaluation of the determinant is again straightforward. In this case, only 
massless RR closed string states contribute to the trace, massive modes cancelling 
by world-sheet supersymmetry. One finds
\be
Z_K = \int\! d^8x_0 \!\int\! d^8\psi_0 \, \prod_{i=1}^4 
\left(\frac{\lambda_i/2\pi}{\tanh \lambda_i/2\pi }
\right) = \int\! d^8x_0\int\! d^8\psi_0 \, \hat {\cal L}(R) 
\label{ZZK}
\ee
As expected, the final result is independent of the modulus $t$. 
In this case, one has to factorize a factor $2^8$ to recover the O8-plane 
charge squared ${\mu_8^\prime}^2 = 256 \mu_8^2$. This amounts to turn each 
$R$ into $R/4$. Extrapolating the results of the previous 
section, it is then natural to expect that this partition function should correspond
to the square of the unknown O8-plane gravitational coupling, so that the analog
of (\ref{constrainta}) and (\ref{constraintm}) is
\be 
\hat {\cal O} ({\cal R}) \wedge \hat {\cal O} ({\cal R}) = \hat {\cal L}({\cal R}/4)
\label{constraintk}
\ee
which is satisfied for all the components by the solution (\ref{sol}).

The subtle point in this derivation is that, in writing the light-cone trace 
(\ref{ZK}), we completely neglected the ghost and superghost contributions as well
as the bosons and fermions in the two remaining flat directions. It is actually a subtle
issue to give a meaning to these contributions; luckily, we are interested on orientifold 
couplings only, that are completely encoded in (\ref{ZK}) alone, in our kinematical set up.
This means that the possible ambiguity of the full result,
due to the issues above, will only concern the nature of the RR propagator flowing in the
surface. Anyway, despite the extrapolation needed to obtain an information from the
Klein bottle amplitude, the compatibility with the couplings obtained from the annulus 
and the M\"obius strip constitute a non-trivial consistency check of the whole approach
we have followed.


\section{Normal bundle}
\setcounter{equation}{0}

Magnetic interactions among D8-branes and O8-planes are 
not suitable to extract the anomalous couplings coming from the expansion of the 
normal bundle. On the other hand, although the RR odd spin structure amplitudes 
between parallel Dp-branes and/or Op-planes with $p\leq 7$ are necessarily vanishing
due to a simple counting of fermionic zero modes, this does not mean that one can not 
extract information from these.
In particular, thanks to (\ref{propodd}), it is clear what is the reason why the
amplitude is vanishing; for $p\leq 7$, the presence of more than one fermionic
zero mode in the Dirichlet directions (one can always be absorbed in the 
$\partial_{P}\Delta$ of (\ref{propodd})) unavoidably
makes the correlator (\ref{propodd}) vanish. A more useful way of writing
the propagator (\ref{propodd}) is
\be
\langle C_{(p)}^{M_1 ... M_p}  H_{(q+1)}^{N_1 ... N_{q+1}} \rangle_{odd} 
\sim \,\int\!(\prod_{i=1}^{10}d\psi_0^{M_i})\psi_0^{M_1}...
\psi_0^{M_p}\psi_0^{N_1}...\psi_0^{N_{q+1}}\psi_0^P
\partial_{P}\Delta
\label{propodd2}
\ee
A generic amplitude ${\cal M}$ in the RR odd spin structure can be written as 
\be
{\cal M}=\int\!(\prod_{i=1}^{10}d\psi_0^{M_i}) \tilde{{\cal M}}
\label{M0}
\ee
displaying explicitly that ${\cal M}$ vanishes unless $\tilde{{\cal M}}$
does not provide the ten fermionic zero modes $\psi_0^{M}$.
As we mentioned, the amplitude ${\cal M}$ will in general encode not only 
information about the couplings (\ref{WZD}), (\ref{WZO}), but also possible
corrections to the free propagator (\ref{propodd2}). 

\subsection{Annulus and M\"obius strip}

\vskip 2pt

In order to understand how one can distinguish between these two different 
contributions, let us focus on the particular case we will be interested
in the following, i.e. the case in which ${\cal M}$ is the correlation function
of one world-volume gauge field with two gravitons in the bulk for parallel 
D4-branes/O4-planes\footnote{This is not an essential condition but
it simplifies the analysis below.}.
At the level of the world-volume effective action, these interactions are then given
by a tree level amplitude of 2 gravitons and one gauge field, mediated by the
propagator (\ref{propodd2}) above, see figure 3.2. 
Doing the same manipulations as in previous 
section, the part of the unknown D4-brane Wess-Zumino coupling linear in the gauge 
field $A$ can be written as
\be
S_{D4} (A) = \mu_4 \int \left. A \wedge H \wedge \hat {\cal D}({\cal R}, {\cal R^\prime}) 
\right|_{5-form} \label{CD4A}
\ee
where ${\cal R}, {\cal R}^\prime$ are the normalized tangent and normal bundle curvature two-forms.
The other source is now  either a D4-brane without gauge field or a O4-plane,
with Wess-Zumino couplings
\be
S_{D4} = \mu_4 \int \left. C \wedge \hat {\cal D}({\cal R}, {\cal R}^\prime) \right|_{5-form} \;,\;\;
S_{O4} = \mu^\prime_4 \int \left. C \wedge \hat {\cal O}({\cal R}, {\cal R}^\prime) \right|_{5-form}
\label{CD4O4}
\ee
It is then easily seen that there are no non-vanishing 
overlaps of RR forms through the propagator (\ref{propodd2}), since 
always at least four fermionic zero modes are missing and make 
(\ref{propodd2}) vanish.
By comparing the form of the lagrangian (\ref{CD4A}) to the free propagator
(\ref{propodd2}) and considering that the derivative acting on $\Delta$ has
to be along a Dirichlet coordinate, we may easily conclude that the couplings in
(\ref{CD4A}) are encoded in the part of the amplitude $\tilde{{\cal M}}$
containing five zero modes in Neumann directions and one along a Dirichlet direction,
which can be easily selected by inserting by hand the remaining four transverse 
(properly normalized) fermionic zero modes $(\sqrt{t}\psi_0^5) ...(\sqrt{t}\psi_0^8)$. 
All other contributions in 
$\tilde{{\cal M}}$ should be understood as corrections to the free propagator 
in (\ref{propodd}) or eventually as new kind of couplings, not appearing in 
(\ref{CD4A}). Although the study of such contributions is definitively interesting, 
for the rest of the paper our attention will be focused on the anomalous couplings 
(\ref{WZD}), (\ref{WZO}) only. 

\begin{figure}[h]
\vskip 10pt
\centerline{$\raisebox{22pt}{$a)$}$ \epsfysize = 2cm \epsffile{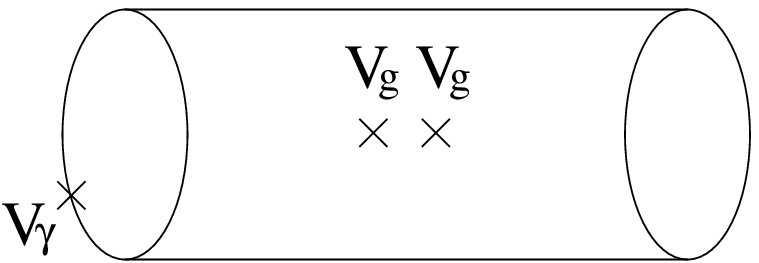}}
\vskip 20pt
\centerline{$\raisebox{22pt}{$b)$}$ \epsfysize = 2cm \epsffile{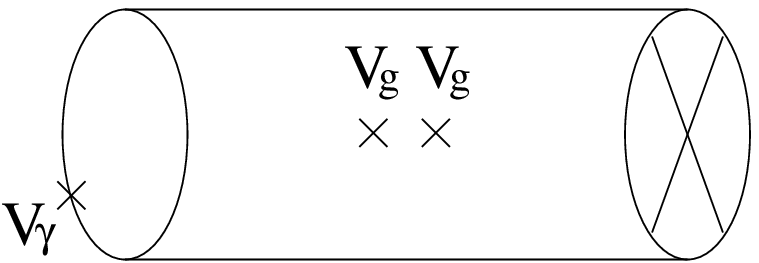}}
\vskip 10pt
\centerline{Fig. 3.1: {\it The correlation $I_{\gamma g^2}$ on a) the annulus and b) the 
M\"obius strip.}}
\vskip 10pt
\end{figure}

We have then to evaluate (taking again the photon vertex in the ($-1$)-picture) the
correlation
\be
I_{\gamma g^2} = \langle t^2\,\psi_0^5 \psi_0^6 \psi_0^7 \psi_0^8 \,
(T_F + \tilde T_F) V_\gamma^{(-1)}\,V_g^{(0)}\,V_g^{(0)}\,
\rangle
\label{2g} 
\ee
on the annulus and M\"obius strip. Similarly to the computation of last section,
it is sufficient to consider the linearized part of the vertices in (\ref{2g}),
having in mind that gravitons carry now generic polarizations and momenta. 
It is clear that the two graviton vertices have to provide
at least four zero modes, in order to make (\ref{2g}) non-vanishing. 
The linearized graviton vertex (\ref{vgravlin}) decomposes in this case into various 
pieces, according to the number of fermionic zero modes provided by each term.
Since we consider trivial embeddings of D-branes (O-planes) in ten-dimensional space-time,
the graviton vertices have to provide fermionic zero modes in the Neumann directions
only. By using again the cyclic property of the Riemann tensor (\ref{RT}), 
it is straightforward to see that the effective graviton operator is now
\bea
\tilde{V}_g^{eff.}\a=\a \int \! d^2z\, \left\{R_{\mu\nu}(p)\,\left[
X^\mu(\partial+\bar{\partial})X^\nu+
(\psi-\tilde{\psi})^\mu (\psi-\tilde{\psi})^\nu \right]
\right. \nn \\ \a\;\a \hspace{38pt} 
\left. +R^\prime_{ij}(p) \, \left[X^i(\partial+\bar{\partial})X^j+
(\psi-\tilde{\psi})^i (\psi-\tilde{\psi})^j \right] \right\}
\label{veff2}
\eea
involving both the tangent and normal bundle $SO(4)$-valued curvature two-forms
\be
R_{\mu\nu} = \frac 12 R_{\mu\nu\rho\sigma} \psi_0^\rho \psi_0^\sigma \;,\;\;
R^\prime_{ij} = \frac 12 R_{ij\rho\sigma} \psi_0^\rho \psi_0^\sigma
\label{tncurv}
\ee
The remaining Neumann and Dirichlet
zero modes in the $0,9$ directions will be provided respectively by the photon 
vertex and the picture changing operator, precisely like in the previous case. 
The correlator (\ref{2g}) then reduces to 
\be
I_{\gamma g^2} = T \int_0^\infty \! \frac {dt}t \,(2 \pi t)^{-1/2} 
\,e^{- b^2 t / (2 \pi^2)} \, A_0 \, bt^3 \, I^{eff.}_{g^2}
\ee
in terms of the effective two-point function
\be
I^{eff.}_{g^2} = \langle \langle \tilde{V}_g^{eff.}\,
\tilde{V}_g^{eff.}\,\rangle \rangle
\label{2geff}
\ee
in four space-like world-volume directions and four transverse directions. 
One can again exponentiate the correlation function (\ref{2geff}), reducing the
computation to the evaluation of the determinant of the same twisted action
(\ref{Sint}), but with the appearence of the curvature of both the tangent and 
normal bundles.
The computation of the determinants is similar to the tangent bundle case. 
As shown in section four, the net result is that the determinant for Dirichlet 
directions is exactly the inverse of the Neumann one and therefore one gets
\bea
Z_{A,M}(t) \a=\a \int\! d^4 x_0 \!\int\! d^4 \psi_0 \, 
\prod_{i=1}^2 \left(\frac{\lambda_i/4 \pi}{\sinh \lambda_i/4 \pi}\right) 
\prod_{j=1}^2 \left(\frac{\sinh \lambda^\prime_j/4 \pi}{\lambda^\prime_j/4 \pi}\right)
\nn \\ \a=\a \int\! d^4 x_0 \int\! d^4 \psi_0 \, \hat {\cal A}(R) / \hat {\cal A}(R^\prime)  
\label{ZZAM'}
\eea
where $\lambda_i$ and $\lambda^\prime_i$ are the skew-eigenvalues of $R_{\mu\nu}$ and
$R^\prime_{ij}$ respectively.
Finally, the conveniently normalized results 
for the annulus and M\"obius strip are
\bea
I^A_{\gamma g^2} \a=\a T \, \mu_4^2 \int \!d^4x_0 \, \epsilon_{\mu_1 ... \mu_{5}} 
\left(\hat {\cal A}({\cal R}) / \hat {\cal A}({\cal R}^\prime) \right)_4^{\mu_1 ... \mu_4}
A^{\mu_5} \partial \Delta_{(5)} (b) 
\label{IA'}\\
I^M_{\gamma g^2} \a=\a T \, \mu_4 \mu_4^\prime \int \!d^4x_0 \, \epsilon_{\mu_1 ... \mu_5} 
\left(\hat {\cal A}({\cal R}/2) / \hat {\cal A}({\cal R}^\prime/2) \right)_4^{\mu_1 ... \mu_4} 
A^{\mu_5} \partial \Delta_{(5)} (b)
\label{IM'}
\eea
where $\Delta_{(5)}(b)$ is the scalar Green function in the five transverse dimensions.

\begin{figure}[h]
\vskip 10pt
\centerline{$\raisebox{22pt}{$a)$}$ \epsfysize = 2cm \epsffile{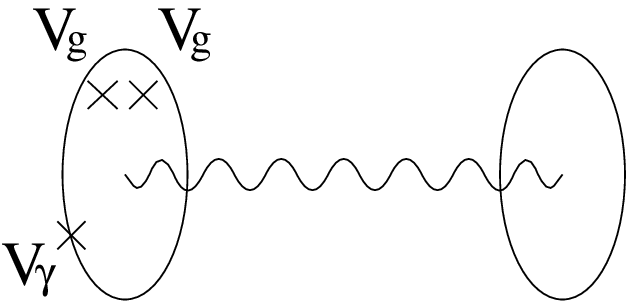} 
$\raisebox{22pt}{$+\!\!$}$ \epsfysize = 2cm \epsffile{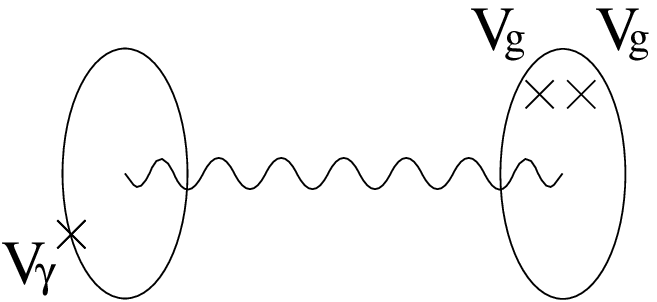}}
\vskip 20pt
\centerline{$\raisebox{22pt}{$b)$}$ \epsfysize = 2cm \epsffile{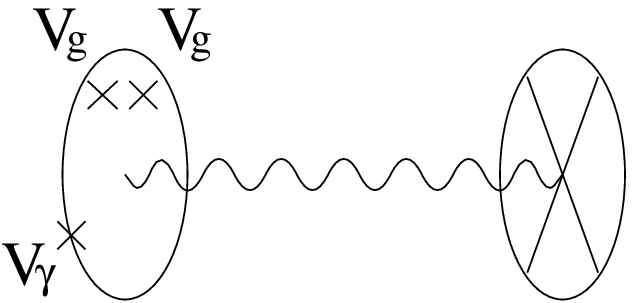} 
$\raisebox{22pt}{$+\!\!$}$ \epsfysize = 2cm \epsffile{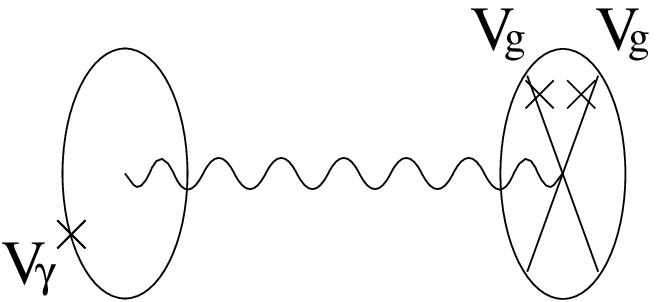}}
\vskip 10pt
\centerline{Fig. 3.2: {\it Factorization of $I_{\gamma g^2}$ on a) the annulus 
and b) the M\"obius strip.}} 
\vskip 10pt
\end{figure}

Again, it is straightforward to deduce the gravitational anomalous couplings 
(\ref{WZD}) and (\ref{WZO}) by comparing (\ref{IA'}) and (\ref{IM'}) to 
the expressions one gets by using the generic couplings $\hat {\cal D}({\cal R}, {\cal R}^\prime)$
and $\hat {\cal O}({\cal R}, {\cal R}^\prime)$, and the propagator (\ref{propodd2}) (modified by
the insertion of the four transverse zero modes), which are
\bea
\hspace{-10pt} I_{D4-D4} \a=\a T \, \mu_4^2 \int \!d^4x_0 \, \epsilon_{\mu_1 ... \mu_5} 
\left(\hat {\cal D} ({\cal R}, {\cal R}^\prime) \wedge \hat {\cal D} ({\cal R}, {\cal R}^\prime) 
\right)_4^{\mu_1 ... \mu_4} A^{\mu_5} \partial \Delta_{(5)} (b) 
\label{ID4D4'} \\ \hspace{-10pt}
I_{D4-O4} \a=\a T \, \mu_4 \mu_4^\prime \int \!d^4x_0 \, \epsilon_{\mu_1 ... \mu_5} 
\left(\hat {\cal D} ({\cal R}, {\cal R}^\prime) \wedge \hat {\cal O}({\cal R}, {\cal R}^\prime) 
\right)_4^{\mu_1 ... \mu_4} A^{\mu_5} \partial \Delta_{(5)} (b)
\label{ID4O4'}
\eea
By comparison one finds two conditions generalizing the conditions 
(\ref{constrainta}) and (\ref{constraintm}), which have the unique solutions
\be
\hat {\cal D} ({\cal R}, {\cal R}^\prime) = \sqrt{\hat {\cal A} ({\cal R}) / 
\hat {\cal A} ({\cal R}^\prime)} \;,\;\; 
\hat {\cal O} ({\cal R}, {\cal R}^\prime) = \sqrt{\hat {\cal L} ({\cal R}/4) / 
\hat {\cal L} ({\cal R}^\prime/4)} 
\label{sol'}
\ee

\subsection{Klein bottle}

\vskip 2pt

Again, the Klein bottle computation related to the O4-O4 interaction can not be done 
along the same lines, since it is not possible to insert any photon. 
Nevertheless, one can observe as in the case of the tangent bundle, that the effective 
light-cone partition function makes sense and can be computed. 
Similarly to the previous cases, one finds
\bea
Z_K \a=\a \int\! d^4 x_0 \!\int\! d^4 \psi_0 \, 
\prod_{i=1}^2 \left(\frac{\lambda_i/2 \pi}{\tanh \lambda_i/2 \pi}\right) 
\prod_{j=1}^2 \left(\frac{\tanh \lambda^\prime_j/2 \pi}{\lambda^\prime_j/2 \pi}\right)
\nn \\ \a=\a \int\! d^4 x_0 \!\int\! d^4 \psi_0 \, \hat {\cal L}(R) / \hat {\cal L}(R^\prime)  
\label{ZZK'}
\eea 
This result suggests a condition analog to (\ref{constraintk}), which is compatible with 
the solution (\ref{sol'}).
As for the case of the tangent bundle, the significance of the Klein bottle computation
lies in its compatibility with the more rigorous results derived from the annulus and
M\"obius strip.


\section{$\sigma$-model approach}
\setcounter{equation}{0}

As seen in last two sections, the couplings (\ref{WZD}) and (\ref{WZO}) are
basically encoded in an eight-dimensional light-cone partition function on various 
surfaces. More precisely, according to the discussion of last section, 
these couplings are determined by terms involving only fermionic zero modes along 
the space-like directions of the world-volume of the two parallel sources. 
This allows one to consider the indices (\ref{ZAM}) and (\ref{ZK}) in the 
more general case of interactions between Dp-branes and/or Op-planes for generic p.
For completeness, we also include in the following the dependence on a constant
${\cal F} = 2 \pi \al F - B$\footnote{In this and the following sections, we shall keep track
of all the normalizations and restore the $\al$ dependence.}.

The path integral representation for the indices (\ref{ZAM}) and (\ref{ZK}) is
\be
Z_\Sigma=\int_{\Sigma}{\cal D}X{\cal D}\Psi\,e^{-S(X,\Psi)}
\label{path}
\ee
where $\Sigma=A,M,K$ denotes the corresponding surface, annulus, M\"obius strip and
Klein bottle, with the appropriate boundary conditions for all the fields.
$S(X,\Psi)$ is the (Euclidean) two-dimensional supersymmetric $\sigma$-model defined
on an eight-dimensional target space\footnote{As explained in \cite{wittb}, the net effect of
a constant antisymmetric NSNS tensor $B_{\mu\nu}$ is just to shift 
$F_{\mu\nu} \rightarrow {\cal F}_{\mu\nu} / 2 \pi \al$.}:
\bea
S \a\!=\a\! \frac 1{4 \pi \al} \! \int_\Sigma \! d^2\sigma 
\! \left[g_{MN}\left(\partial^\alpha X^M\partial_\alpha X^N + 
2 \al \bar{\Psi}^M \hat D \Psi^N \right) 
+ \frac{\!\!(2\al)^2\!\!\!}{6} R_{MNPQ}  (\bar{\Psi}^M\Psi^P)(\bar{\Psi}^N\Psi^Q) \right] 
\nn \\ \a\;\a + \frac 1{4 \pi \al} \oint_{\partial \Sigma} \!\! d \sigma_\alpha {\cal F}_{\mu\nu} 
(X^\mu\partial^\alpha X^\nu + 2 \al \bar{\Psi}^\mu\rho^\alpha\Psi^\nu)
\label{actionsigma}
\eea
where the $U(1)$ field strength ${\cal F}_{\mu\nu}$ is constant.
Capital indices $M,N=1,..., 8$ run now over $\mu,\nu=1,...p$ (Neumann)
and $i,j=p+1,...8$ (Dirichlet). In the action (\ref{actionsigma}), 
$\Psi = (\psi, \tilde{\psi})$, $\hat \partial = \rho^\alpha \partial_\alpha$ and
\be
\hat D (X) \Psi^N = \hat \partial \Psi^N +  
\Gamma^N_{PQ}(X)(\hat \partial X^P)\Psi^Q
\ee
The world-sheet is parametrized by $\sigma_\alpha=\sigma_0,\sigma_1$,
with $0\leq \sigma_0\leq t$, $0\leq\sigma_1\leq \pi$, and
\be
\rho^0 = \pmatrix{ 0 \a -i \cr i \a 0\cr} \;,\;\;  
\rho^1 = \pmatrix{ 0 \a 1 \cr 1 \a 0 \cr} \;,\;\; \rho^3 = \rho^0 \rho^1
\ee
Since all the partition functions that we are computing are indices, independent of
$t$, the path integral (\ref{path}) can be evaluated in the high temperature regime
$t\rightarrow 0$ \cite{agw}. In this limit, the path-integral is dominated by 
constant paths and it is sufficient to consider quadratic fluctuations around 
this configuration. In all the surfaces involved, bosons present constant 
configurations $x_0^{\mu}$ in the Neumann directions only,
whereas fermions always have the constant modes $\psi_0^\mu=\tilde{\psi}_0^\mu$, 
$\psi_0^i=-\tilde{\psi}_0^i$.
Using the normal coordinate expansion \cite{agfm}, one can write the following
expansions for the metric and the connection
\bea
g_{MN}(X)\a=\a \delta_{MN}-\frac 13 R_{MPNQ}(x_0)\xi^P\,\xi^{Q} + ... \nn \\
g_{QP}(X) \, \Gamma_{MN}^{P}(X) \a=\a \frac 13
(R_{QMPN}(x_0)+ R_{QNPM}(x_0))\, 
\xi^P + ...
\label{norm}
\eea
where $\xi^M$ are the fluctuations of the bosonic fields. 
The bosonic kinetic term gives simply
\be
g_{MN}(X)\,\partial^\alpha X^M \partial_\alpha X^N =
\partial^\alpha \xi^M \, \partial_\alpha \xi_M + ...
\label{kinbos}
\ee
As we have explained, only terms bilinear in the Neumann fermionic zero modes
contribute to the couplings we are interested in. Restricting then to those
terms only, using the cyclic property of the Riemann tensor and the boundary 
condition $\psi_0^\mu=\tilde{\psi}_0^\mu$, the fermionic kinetic term yields
\be
g_{MN}(X)(\bar{\Psi}^M \hat D \Psi^N)=
\bar{\chi}^M \hat \partial \chi_{M}
+R_{\mu\nu MN} (x_0) \,\psi_0^\mu\,\psi_0^\nu\,\xi^M \,\partial_0 \xi^N + ...
\label{kinfer2}
\ee
where $\chi$ represents the fluctuation of two dimensional spinor field.
The four-fermion term in (\ref{actionsigma}) gives, after some simple manipulations,
\be
R_{MNPQ}(X)(\bar{\Psi}^M \Psi^P)(\bar{\Psi}^N \Psi^Q)=
3\, R_{\mu\nu MN} (x_0) \,\psi_0^\mu\,\psi_0^\nu \, (\tilde{\chi}-\chi)^M
(\tilde{\chi}-\chi)^N+ ...
\label{kinfer}
\ee
Putting (\ref{kinbos}), (\ref{kinfer2}) and (\ref{kinfer}) together,
one finds the usual free action
\be
S_0 = \frac 1{4 \pi \al} \int d^2\sigma\, \left[ (\partial_\alpha\xi^M)^2 
+ 2 \al \bar{\chi}^M \hat \partial \, \chi_{M} \right] 
\ee
and the gravitational interaction term
\bea
S^{curv}_{int} \a=\a \frac 1{4 \pi}\int \! d^2z\, \left\{R_{\mu\nu}\,\left[
\xi^\mu(\partial+\bar{\partial})\xi^\nu + 2 \al (\chi-\tilde{\chi})^\mu 
(\chi-\tilde{\chi})^\nu \right]\right. \nn \\ \a\;\a \hspace{50pt} 
\left. + R^\prime_{ij} \, \left[\xi^i(\partial+\bar{\partial})\xi^j + 2 \al
(\chi-\tilde{\chi})^i (\chi-\tilde{\chi})^j \right] \right\}
\label{scurv}
\eea
where $R_{\mu\nu}$ and $R^\prime_{ij}$ are defined as in (\ref{tncurv}).

The last term in the action (\ref{actionsigma}) leads to an effective boundary interaction 
lagrangian (keeping only bilinears in the fermionic zero modes)
\be
S^{gauge}_{int}= \frac 1{\pi} \oint \! ds \,
{\cal F}_{\mu\nu} \, \psi_0^\mu \psi^\nu_0
\label{sgauge}
\ee 
As expected, the action (\ref{scurv}) reproduces precisely the interactions
obtained in last sections by exponentiating the effective graviton vertices (\ref{veff})
and (\ref{veff2}), replacing $X$ and $\psi$ by the quantum fields $\xi$ and $\chi$ 
respectively. The effective $\sigma$-model interaction lagrangian is finally given by 
$S=S_0+S^{curv}_{int}+S^{gauge}_{int}$.
By an orthogonal transformation we can always bring $R_{MN}$ in a ``skew-diagonal'' 
form
\bea
R_{MN}=\pmatrix{R_{\mu\nu}\a \cr \a R_{ij}^\prime}=
               \pmatrix {0 \a \lambda_1 \a \a \a \a \cr
                     -\lambda_1 \a 0 \a \a \a \a \cr
                      \a \a ... \a \a \a \cr
                      \a \a \a 0 \a \lambda^\prime_1 \a \cr
                      \a \a \a -\lambda^\prime_1 \a 0 \a \cr
                      \a \a \a \a \a ... \cr}
\label{matrix}
\eea
The partition function $Z$ decouples then into four different pieces,
according to the decomposition given by (\ref{matrix}). The evaluation of
the determinant for each pair of coordinates is straightforward and can be
performed by expanding in modes the boson and fermion fields on the 
various surfaces (see the appendix). Only the constant mode in the $\sigma$ direction 
contribute to the trace, all other modes cancelling by world-sheet supersymmetry.  

On the annulus and the M\"obius strip, the fermionic term in the tangent 
bundle contribution to (\ref{scurv}) vanishes, due to Neumann boundary conditions,
so that the path integral over 
$\chi^\mu$ gives no net contribution. The path-integral for $\xi^\mu$ is
instead that of a scalar in the $R_{\mu\nu}$ background that is equivalent to a 
constant electromagnetic field.
On the other hand, as for the normal bundle contribution, where the fields satisfy
Dirichlet boundary conditions, there are no bosonic $\sigma$-independent modes on 
both surfaces, as is clear from (\ref{Amode}) and (\ref{Mmode}). However, due to 
the different sign in the boundary conditions (\ref{bcAD}) and
(\ref{bcMD}), the fermionic contributions do not cancel anymore from (\ref{scurv}), 
as happens for the tangent bundle case. Including also the boundary term (\ref{sgauge}),
one finally finds 
\bea
Z_{A,M} \a=\a \int\! d^p x_0 \!\int\! d^p \psi_0 \, (4 \pi \al t)^{- \frac p2} 
\, e^{\frac {\cal F}{\pi}} 
\prod_{i=1}^{q} \left(\frac{\al \lambda_i t}{\sinh \al \lambda_i t}\right) 
\prod_{j=1}^{4-q} \left(\frac{\sinh \al \lambda^\prime_j t}{\al \lambda^\prime_j t}\right) \nn \\
\a=\a \int\! d^p x_0 \!\int\! d^p \psi_0 \, e^{\frac {\cal F}{4 \pi^2 \al}} \prod_{i=1}^{q}
\left(\frac{\lambda_i/4\pi}{\sinh \lambda_i/4\pi}\right) 
\prod_{j=1}^{4-q} \left(\frac{\sinh \lambda^\prime_j/4\pi}
{\lambda^\prime_j/4\pi}\right) \nn \\
\a=\a \int\! d^p x_0 \! \int\! d^p \psi_0 \,e^{\frac {\cal F}{4 \pi^2 \al}} 
\hat {\cal A}(R) / \hat {\cal A}(R^\prime)  
\label{ZZAMs}
\eea
where ${\cal F} = {\cal F}_{\mu\nu}\psi_0^\mu \psi_0^\nu$ (for the annulus, it is 
just the difference of the field strengths on the two boundaries) and
$2q<p$ is the dimension of the curved submanifold of the Dp or Op world-volume
(with non-trivial characteristic classes defined by the roof genus or Hirzebruch
polynomials)\footnote{Notice that (\ref{ZZAMs}) and (\ref{ZZKs}) are
polynomials in the curvature two-forms and therefore 
should be always integrated over a manifold with an even dimension $2q$.
Moreover, the polynomials defined by these expressions are given always
by even powers of $\lambda_i$ (quadratic in the fermionic zero modes) and
therefore we need at least a 4-dimensional manifold to get a non-trivial
result.}. In the second line of (\ref{ZZAMs}) (and similarly for (\ref{ZZKs})),
we have used the fact that the integral over the $p$ fermionic zero modes selects
only the p-form part of the integrand, which scales like $\prod_{i=1}^q \lambda_i$. 
One can therefore drop the free particle normalization $(4 \pi \al t)^{-p/2}$ and turn
each $\al \lambda_i t$ into $\lambda_i / 4 \pi$. 

On the Klein bottle, only RR massless closed string states contribute,  
corresponding again to world-sheet fields which are constant in $\sigma$. 
In this case, boson and fermion fields contribute each to both the tangent and the normal
bundle contributions, with their role exchanged in the two cases because of the different 
boundary conditions, so that the result for Dirichlet directions is the inverse of 
that for Neumann directions. One then finds
\bea
Z_K \a=\a \int\! d^p x_0 \int\! d^p \psi_0 \, (4 \pi \al t)^{- \frac p2}
\prod_{i=1}^{q} \left(\frac{2 \al \lambda_i t}{\tanh 2 \al \lambda_i t}\right) 
\prod_{j=1}^{4-q} \left(\frac{\tanh 2 \al \lambda^\prime_j t}{2 \al \lambda^\prime_j t}\right) 
\nn \\ \a=\a \int\! d^p x_0 \int\! d^p \psi_0 \, 
\prod_{i=1}^{q} \left(\frac{\lambda_i/2\pi}{\tanh \lambda_i/2\pi}\right) 
\prod_{j=1}^{4-q} \left(\frac{\tanh \lambda^\prime_j/2\pi}{\lambda^\prime_j/2\pi}\right) 
\nn \\ \a=\a \int\! d^p x_0 \int\! d^p 
\psi_0 \, \hat {\cal L}(R) / \hat {\cal L}(R^\prime)  
\label{ZZKs}
\eea 

Notice that the partition functions (\ref{ZZAMs}) and (\ref{ZZKs}) are integer numbers
and do not depend on $t$ and $\al$, as expected. By factorizing in both (\ref{ZZAMs}), 
(\ref{ZZKs}) the appropriate factors needed to reconstruct the charges of the two sources, 
equations (\ref{ZZAMs}) and (\ref{ZZKs}) reproduce in particular
(\ref{ZZAM}), (\ref{ZZK}), (\ref{ZZAM'}) and (\ref{ZZK'}).


\section{Consistency with anomaly cancellation in Type I theory}
\setcounter{equation}{0}

As mentioned in the introduction, in this section we will perform an important check 
on the normalization of our results by using the known results for anomaly cancellation
in the Type I superstring. 
It is well known that gauge and gravitational hexagon anomalies in Type I 
theory cancel against certain anomalous diagrams involving the 
exchange of the RR 2-form $C_{(2)}$, if the gauge group is $SO(32)$ \cite{GS}. 
In a modern language in which Type I string theory is
considered as a theory of 32 overlapping D9-branes and one O9-plane
(realizing the world-sheet parity projection), the Green-Schwarz anomaly cancelling 
terms are encoded in the Wess-Zumino couplings (\ref{WZD}), (\ref{WZO}).
In particular the sum of the O-plane and D-brane couplings has to match up in
\be
S_{GS} = \mu_9 \int \left((\pi \al)^2 C_{(6)} \wedge X_4 + (\pi \al)^4 C_{(2)}
\wedge X_8 \right)
\label{WZGS}
\ee
where 
\bea
X_4 \a=\a 2 \,\mbox{tr} F^2 - 2 \,\mbox{tr} R^2 \\
X_8 \a=\a \frac 23 \,\mbox{tr} F^4 - \frac 1{12} \,\mbox{tr} F^2 \, \mbox{tr} R^2 +
\frac 1{12} \,\mbox{tr} R^4 + \frac 1{48} \, (\mbox{tr} R^2)^2
\eea
and the trace is taken in the fundamental representation of the gauge group $SO(32)$.
The first term in (\ref{WZGS}) is required to get the correct Bianchi identity for 
the field strength associated to $C_{(2)}$ 
\be
dH_{(3)}=\frac{1}{16\pi^2}( \mbox{tr} R^2-\mbox{tr} F^2 )
\ee
whereas the second one reproduces the usual anomalous coupling of the 
antisymmetric tensor $C_{(2)}$. Using the explicit expressions 
\be
p_1(R) = \frac 1{2(2 \pi)^2} \, \mbox{tr}R^2 \;,\;\;
p_2(R) = - \frac 1{4(2 \pi)^4} \left(\mbox{tr}R^4 - \frac 12 (\mbox{tr} R^2)^2 \right)
\ee
for the first two Pontrjagin classes $p_1(R),p_2(R)$, the anomalous couplings 
deduced in last section can be rewritten for the D9-branes and O9-planes  
as
\bea
S_{D9} = \mu_9 \int \left(C_{(10)} + (\pi \al)^2 C_{(6)} \wedge B_4 + 
(\pi \al)^4 C_{(2)} \wedge B_8 \right)
\label{WZD9}
\eea
with
\bea
B_4 \a=\a 2 \, F^2 - \frac 1{24} \,\mbox{tr} R^2 \\
B_8 \a=\a \frac 23 \, F^4 - \frac 1{12} \, F^2 \,\mbox{tr} R^2 + \frac 1{720} \,\mbox{tr} R^4
+ \frac 1{1152} \, (\mbox{tr} R^2)^2
\eea
and
\bea
S_{O9} = - \mu_9 \int \left(32 C_{(10)} + (\pi \al)^2 C_{(6)} \wedge O_4 + 
(\pi \al)^4 C_{(2)} \wedge O_8 \right)
\label{WZO9}
\eea
with
\bea
O_4 \a=\a \frac 23 \,\mbox{tr} R^2 \\
O_8 \a=\a - \frac 7{180} \,\mbox{tr} R^4 + \frac 1{144} \,(\mbox{tr} R^2)^2
\eea
One can then check that\footnote{The cancellation of the total charge with respect 
to the ten form $C_{(10)}$ is the notorious tadpole cancellation for the gauge group 
SO(32).}
\be
S_{32 D9}+S_{O9}=S_{GS}
\label{TypeI}
\ee
Any change in the normalization of (\ref{WZD}) and (\ref{WZO}) would invalidate
this result. In particular, assume only the validity of the functional dependence 
of the couplings (\ref{WZD}) and (\ref{WZO}), but take generically
$\hat {\cal A}(c_D \pi^2 \al R)$ in (\ref{WZD}) and $\hat {\cal L}(c_O \pi^2 \al R)$ 
in (\ref{WZO}), where $c_D$ and $c_O$ are arbitrary coefficients. Imposing then 
(\ref{TypeI}), one finds the unique solution $c_D=4,c_O=1$.
Note moreover that the matching with Type I anomaly cancellation (\ref{TypeI})
yields more than two constraints for the parameters $c_D,c_O$. This means that
already the existence of the solution above is a strong 
consistency check of the correctness of the couplings we find.

As anticipated in the introduction, the constraint (\ref{TypeI}) allows to fix only 
the normalizations associated to the roof genus and Hirzebruch polynomials of the tangent 
bundle only. It is however clear from our one-loop correlation functions that no extra numerical 
factors enter into the evaluation of the anomalous couplings coming from the roof 
genus and Hirzebruch polynomials associated to the normal bundle. 


\section{Discussion and conclusions}
\setcounter{equation}{0}

It is quite interesting to analyze more closely the kind of anomalies
computed in last sections, by looking at the string states propagating
in the various surfaces. For the case of D8-D8 interactions, the odd spin structure
would suggest that in the open channel one has the propagation of the
parity-odd part of a chiral fermion state. Here parity means ten-dimensional
parity due to the ten fermionic zero modes that give a factor
$\hat{\Gamma}_{11}=\Gamma_0...\Gamma_9$. However, the insertion of a photon
vertex and a picture changing operator in the amplitude, soaking up two zero modes,
turns $\hat{\Gamma}_{11}$ into $\hat{\Gamma}_{9}=\Gamma_1...\Gamma_8$.
This implies that in the effective correlator (\ref{4geff}) or partition function (\ref{Z}),
the open string running in the loop is the parity-odd part of a chiral fermion, but
in the eight-dimensional sense. Since, as we have seen, the correlator (\ref{4geff})
is independent of the modulus $t$ of the annulus, for $t\rightarrow 0$ it will reduce simply
to a loop of an eight-dimensional chiral fermion with four gravitons inserted. 
This is nothing else that the one-loop term giving rise to the gravitational 
contribution to the axial anomaly
in eight dimensions of a chiral fermion. On the other hand, for $t\rightarrow \infty$,
the annulus computation factorizes in the closed string channel to an exchange
of massless closed string states, according to the anomalous couplings (\ref{WZD}).
Being the computation $t$-independent, the two contributions are exactly the same,
providing a very explicit realization of the anomaly inflow mechanism.
The same is valid for the D4-D4 interaction analyzed in section three: 
the zero modes inserted ``by hand'' into the amplitude, in addition to the two provided 
by the photon vertex and the picture changing operator, turn
$\hat{\Gamma}_{11}$ into the four dimensional $\hat{\Gamma}_{5}$. The computation
reproduces then the usual gravitational contribution to the anomaly of an axial current,
due to a chiral fermion \cite{dsef}. The same is valid of course for arbitrary
Dp-Dp interactions. The $\sigma$-model approach discussed in section four, for 
the tangent bundle contributions, displays a very close analogy to 
the evaluation of anomalies performed in section 11 of \cite{agw}. In particular, 
as we have seen, the determinants arising in all previous computations are fixed purely 
by the zero energy states of the string, since all other contributions cancel each other.
This implies that we could simply reduce the (1+1)-dimensional $\sigma$-model to a 
(0+1)-dimensional quantum system;
in this way the computation corresponds precisely to the evaluation of the gravitational
contribution to the axial anomaly for a spin 1/2 state coupled to gravity \cite{agw}.
The normal bundle contributions present instead a new feature; due to the different 
boundary conditions of the world-sheet fields, there are no more zero energy bosons, 
whereas fermions now do couple to the curvature at second order. In this way, one gets 
that the anomaly polynomial in this case is just the inverse of the tangent bundle one.

The M\"obius strip case is very similar to the annulus. Again, we have the propagation of
the parity-odd term of a chiral fermion state and the discussion follows precisely
that given above for the annulus case.
The anomalies arising from the Klein bottle surface are more subtle.
Although we do not have a clear understanding of this case, it is quite evident that the
anomaly in this case is due to chiral RR forms, i.e. forms whose field strenghts
are (anti) self-dual. Indeed, for $t\rightarrow 0$, the anomaly polynomial reproduces
the gravitational anomaly of chiral forms discovered in \cite{agw}.
It is quite interesting that, in the string set-up, such anomalies, both from the tangent
and normal bundle cases, are encoded in the index (\ref{ZK}).

We have presented a direct string computation of anomalous
Wess-Zumino couplings for D-branes and O-planes. In much the same 
way as in Polchinski's computation of the RR charge carried
by a Dp-brane and an Op-plane, we have relied on one-loop amplitudes
which, factorized in the closed string channel, encode magnetic RR 
interactions. The advantadge of this method lies in the fact that 
the normalization of the couplings can be fixed in a straightforward
and unambiguous way, in contrast with a more direct RR tadpole
computation, which would be really awkward to normalize.

The importance of this direct check lies in the fact that anomalous
couplings are required for consistency of superstring theories. Also,
they have a number of important consequences and, due to their topological
nature, they are believed to be exact. For instance, the Green-Schwarz
anomaly cancelling term in Type I theory follows from the anomalous
couplings of the background D9-branes and O9-plane. 
Needless to say, it will be extremely interesting to analyze the implications
of the orientifold couplings (\ref{WZO}) in string compactifications to lower dimensions, 
along the lines of \cite{dasmu}, as already done for instance in \cite{cy} for the D-brane
couplings (\ref{WZD}).
We would like to remark that, among the couplings we discussed in this paper, 
there were other interactions terms, that we neglected and that definitively deserves 
a further study.
We think that the computations presented in this work confirms explicitly the couplings
(\ref{WZD}) and gives a direct evidence for the orientifold couplings (\ref{WZO}).


\vspace{5mm}
\par \noindent {\large \bf Acknowledgments}
\vspace{3mm}

We would like to thank M. Bianchi, M. Bill\`o, R. Dijkgraaf, G. Ferretti, 
E. Gava, K. Narain, S. Theisen and G. Thompson for valuable and interesting 
discussions, and S. Mukhi for a useful e-mail correspondence.
M. S. and C.A. S. are grateful respectively to SISSA and the Spinoza Institute for
hospitality. This work has been partially supported by EEC under TMR
contract ERBFMRX-CT96-0045 and by the 
Nederlandse Organisatie voor Wetenschappelijk Onderzoek (NWO). 


\vspace{5mm}
\renewcommand{\theequation}{A.\arabic{equation}}
\setcounter{equation}{0}
\par \noindent {\large \bf A. Mode expansions}
\vspace{3mm}

We report in this appendix the mode expansions for bosons and fermions on 
the annulus, M\"obius strip and Klein bottle surfaces, that are needed to evaluate the
corresponding one-loop determinants. Although most of the content of this appendix
can be found in \cite{mc}, we prefer for completeness to report here those results,
including also the case of Dirichlet boundary conditions, not included in \cite{mc}.
All the considerations that will follow are done in the odd spin structure.
We solve the boundary and crosscap conditions for bosons and fermions on each surface
by extending these fields to their covering torus \cite{burmor}, obtained by appropriate 
identifications of the fields, as shown in figures A.1 a)-c).

\begin{figure}[h]
\vskip 10pt
\centerline{$\raisebox{55pt}{$\;\;\;a)\;$}$ \epsfysize = 3.75cm \epsffile{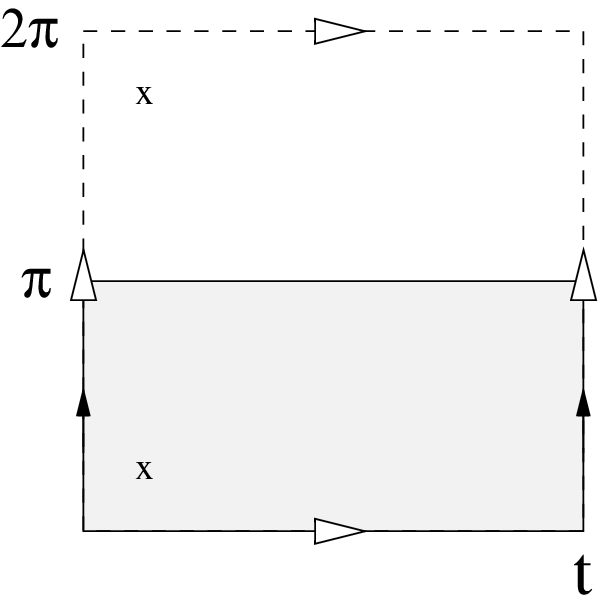}
$\raisebox{55pt}{$\;\;\;b)\;$}$ \epsfysize = 3.75cm \epsffile{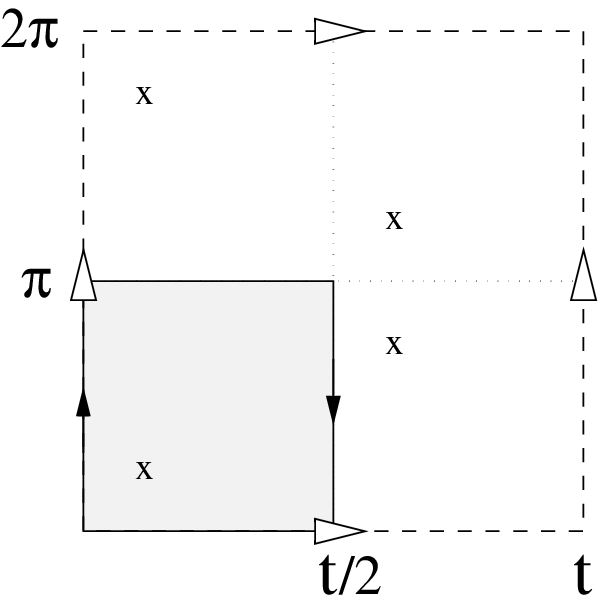}
$\raisebox{55pt}{$\;\;\;c)\;$}$ \epsfysize = 3.75cm \epsffile{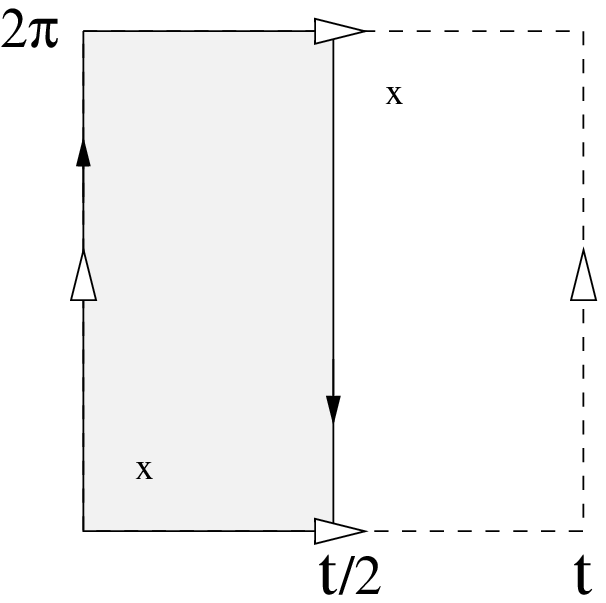}}
\vskip 10pt
\centerline{Fig. A.1: {\it Fundamental cells and covering tori with points identified 
as shown}}
\centerline{{\it for the a) annulus, b) M\"obius strip and c) Klein bottle surfaces.}}
\vskip 10pt
\end{figure}

On the annulus parametrized as in figure A.1 a), the Neumann boundary conditions 
\bea
\partial_{\sigma}X(\tau,0) \a=\a \partial_{\sigma}X(\tau,\pi)=0 \nn \\
\psi(\tau,0) \a=\a \tilde{\psi}(\tau,0) \;,\;\; \psi(\tau,\pi)=\tilde{\psi}(\tau,\pi)
\label{Na}
\eea
and the Dirichlet ones
\bea
X(\tau,0) \a=\a 0 \;,\;\;   X(\tau,\pi)=b \nn \\
\psi(\tau,0) \a=\a -\tilde{\psi}(\tau,0) \;,\;\; \psi(\tau,\pi)=-\tilde{\psi}(\tau,\pi)
\label{bcAD}
\eea
are solved by extending the fields to the covering torus with the identifications 
$X(\tau,\sigma)=X(\tau,2\pi -\sigma)$, $X(\tau,\sigma)=-X(\tau,2\pi -\sigma)+2b$,
respectively for Neumann and Dirichlet boundary conditions, and
$\psi(\tau,\sigma)=\pm\tilde{\psi}(\tau,2\pi -\sigma)$, where here and in the following
the +($-$) sign will always concern the Neumann (Dirichlet) case.
The mode expansions are then:
\begin{eqnarray}
X_{\rm A}^{(N)}(\tau,\sigma)\a=\a\sum_{m=-\infty\atop n\geq0}^{+\infty}\alpha_{m,n}\,
e^{2i\pi m\tau/t}\cos \pi n\sigma \nonumber \\
X_{\rm A}^{(D)}(\tau,\sigma)\a=\a\frac{b}{\pi}\sigma+
\sum_{m=-\infty\atop n>0}^{+\infty}\alpha_{m,n} \,
e^{2i\pi m\tau/t}\sin \pi n\sigma \nn \\
\psi_{\rm A}^{(N,D)}(\tau,\sigma)\a=\a\sum_{m,n=-\infty}^{+\infty}d_{m,n} \, 
e^{2i\pi m\tau/t}e^{i\pi n\sigma} \nn \\
\tilde{\psi}_{\rm A}^{(N,D)}(\tau,\sigma)\a=\a\pm\sum_{m,n=-\infty}^{+\infty}d_{m,n}\, 
e^{2i\pi m\tau/t}e^{-i\pi n\sigma}
\label{Amode}  
\end{eqnarray}
The boundary and cross-cup conditions for the M\"obius strip shown in figure A.1 b) are 
respectively 
\bea
X(0,\sigma)\a=\a X(t/2,\pi -\sigma) \;,\;\;
\partial_{\sigma}X(\tau,0)=\partial_{\sigma}X(\tau,\pi)=0 \nn \\
\psi(0,\sigma)\a=\a\tilde{\psi}(t/2,\pi -\sigma) \;,\;\;
\tilde{\psi}(0,\sigma)=\psi(t/2,\pi -\sigma) 
\eea 
and
\bea
X(0,\sigma)\a=\a-X(t/2,\pi -\sigma) \;,\;\;
X(\tau,0)=X(\tau,\pi)=0 \nn \\
\psi(0,\sigma)\a=\a-\tilde{\psi}(t/2,\pi -\sigma) \;,\;\;
\tilde{\psi}(0,\sigma)=-\psi(t/2,\pi -\sigma) 
\label{bcMD}
\eea 
for the Neumann and Dirichlet cases. The mode expansions on the covering torus are given by
\bea
X_{\rm M}^{(N,D)}(\tau,\sigma)\a=\a\frac 12\sum_{m=-\infty\atop n\geq0}^{+\infty}
\alpha_{m,n}\,e^{2i\pi m\tau/t} (e^{i\pi n\sigma}\pm e^{-i\pi n\sigma}) \;\;,\; 
m+n=even \nonumber \\
\psi_{\rm M}^{(N,D)}(\tau,\sigma)\a=\a\sum_{m,n=-\infty}^{+\infty}d_{m,n}\, 
e^{2i\pi m\tau/t}e^{i\pi n\sigma} \hspace{2.25cm} ,\; m+n=even \nn \\
\tilde{\psi}_{\rm M}^{(N,D)}(\tau,\sigma)\a=\a\pm\sum_{m,n=-\infty}^{+\infty}d_{m,n}\, 
e^{2i\pi m\tau/t}e^{-i\pi n\sigma} \hspace{1.65cm} ,\; m+n=even  
\label{Mmode}
\end{eqnarray}
On the Klein bottle, figure A.1 c), the cross-cup conditions are
\bea
X(\tau,0)\a=\a X(\tau,2\pi) \;,\;\; X(0,\sigma)=\pm X(t/2,2\pi - \sigma) \nn \\
\psi(0,\sigma)\a=\a\pm\tilde{\psi}(t/2,2\pi -\sigma) \;,\;\; 
\tilde{\psi}(0,\sigma)=\pm\psi(t/2,2\pi -\sigma) 
\eea
The corresponding mode expansions are then
\bea
X_{\rm K}^{(N,D)}(\tau,\sigma)\a=\a\frac 12\sum_{m=-\infty\atop n\geq0}^{+\infty}
\alpha_{m,n} \, e^{2i\pi m\tau/t} (e^{i\pi n\sigma}\pm (-)^m e^{-i\pi n\sigma}) 
\nonumber \\
\psi_{\rm K}^{(N,D)}(\tau,\sigma)\a=\a\sum_{m,n=-\infty}^{+\infty}d_{m,n}\, 
e^{2i\pi m\tau/t}e^{i\pi n\sigma} \nn \\
\tilde{\psi}_{\rm K}^{(N,D)}(\tau,\sigma)\a=\a\pm \sum_{m,n=-\infty}^{+\infty}d_{m,n}\, 
(-)^m \, e^{2i\pi m\tau/t}e^{-i\pi n\sigma}
\end{eqnarray}
On all the surfaces the mode expansions for all the fields with Dirichlet boundary
conditions can also be obtained from the mode expansions for the Neumann case by
changing the relative sign between left and right-moving fields. 


\end{document}